\documentclass[manuscript]{aastex631}

\usepackage{CJK}
\usepackage{multirow}
\usepackage{upgreek}

\newcommand{\obj}{SDSS~J111215.82+111745.0}
\newcommand{\jid}{J1112}
\newcommand{\gmd}{$g$-mode}
\newcommand{\pmd}{$p$-mode}
\newcommand{\sn}{Section}
\newcommand{\fn}{Figure}
\newcommand{\tn}{Table}
\newcommand{\an}{Appendix}
\newcommand{\teff}{T_\mathrm{eff}}
\newcommand{\msun}{$M_\sun$}
\newcommand{\rsun}{$R_\sun$}

\newcommand{\uhz}{$\upmu$Hz}
\newcommand{\ud}{\mathrm d}
\newcommand{\mesa}{\tt\string MESA}
\newcommand{\gyre}{\tt\string GYRE}
\newcommand{\extf}{\tt\string run\_star\_extras.f}
\newcommand{\wda}{WD-01789}
\newcommand{\wdb}{WD-02084}

\shorttitle{Asteroseismology of {\obj}}
\shortauthors{Su \& Li}

\begin{document}
\begin{CJK*}{UTF8}{gbsn}

\title{Asteroseismology of the pulsating extremely low-mass white dwarf {\obj}: a model with {\pmd} pulsations consistent with the observations}

\correspondingauthor{Jie Su}
\email{sujie@ynao.ac.cn}

\author[0000-0001-7566-9436]{Jie Su (苏杰)}
\affiliation{Yunnan Observatories, Chinese Academy of Sciences, Kunming 650216, People's Republic of China}
\affiliation{Key Laboratory for the Structure and Evolution of Celestial Objects, Chinese Academy of Sciences, Kunming 650216, People's Republic of China}

\author[0000-0002-1424-3164]{Yan Li (李焱)}
\affiliation{Yunnan Observatories, Chinese Academy of Sciences, Kunming 650216, People's Republic of China}
\affiliation{Key Laboratory for the Structure and Evolution of Celestial Objects, Chinese Academy of Sciences, Kunming 650216, People's Republic of China}
\affiliation{Center for Astronomical Mega-Science, Chinese Academy of Sciences, Beijing 100012, People's Republic of China}
\affiliation{University of Chinese Academy of Sciences, Beijing 100049, People's Republic of China}

\begin{abstract}

{\obj} is the second pulsating extremely low-mass white dwarf discovered. Two short-period pulsations, 107.56 and 134.275\,s, were detected on this star, which would be the first observed pressure mode ({\pmd}) pulsations observed on a white dwarf. While the two potential {\pmd}s have yet to be confirmed, they make {\obj} an interesting object. In this work, we analyzed the whole set of seven periods observed on {\obj}. We adopt three independent period-spacing tests to reveal a roughly 93.4\,s mean period spacing of $\ell=1$ {\gmd}s, which gives added credence to the $\ell=1$ identifications. Then we perform asteroseismic modeling for this star, in which the H chemical profile is taken as a variable. The stellar parameters $M=0.1650\pm0.0137$\,{\msun} and ${\teff}=9750\pm560$\,K are determined from the best-fit model and the H/He chemical profiles are also defined. The two suspected {\pmd}s are also well represented in the best-fit model, and both the stellar parameters and the pulsation frequencies are in good agreement with the values derived from spectroscopy.

\end{abstract}


\section{Introduction} \label{sec:intro}

White dwarf (WD) stars are the final evolutionary stage of the majority of stars in the Galaxy. They thus present an important boundary condition for investigating the previous evolution of stars. In particular, WDs are natural laboratories for studying physical processes under extreme conditions \citep{2008ARA&A..46..157W,2008PASP..120.1043F,2010A&ARv..18..471A}. WDs with $M\lesssim 0.45$\,{\msun} are classified as low-mass WDs, which are thought to have helium cores. There is a subclass of low-mass WDs called extremely low-mass (ELM) WDs. They are characterized by very low masses, i.e. $M\lesssim 0.2$\,{\msun}. It is widely accepted that ELM WDs are originated from the evolution of close binary systems. The interaction between binary stars removes most of the outer envelope of one component before helium ignition. The remnant of the component with a helium core would not go through the asymptotic giant branch phase and would directly contract toward a WD \citep[see][for instance]{2013AA...557A..19A,2016A&A...595A..35I,2019ApJ...871..148L}. 

Some ELM WDs have been detected to exhibit pulsations \citep[for example,][]{2012ApJ...750L..28H,2013ApJ...765..102H,2013MNRAS.436.3573H,2015MNRAS.446L..26K,2015ASPC..493..217B,2017ApJ...835..180B,2018MNRAS.478..867P}. They constitute a separate class of pulsating WDs, commonly referred to as extremely low-mass variables (ELMVs). Pulsations observed in ELMVs are non-radial {\gmd}s (gravity modes), which are excited by the same mechanism as the classical ZZ Ceti stars, that a combination of the $\kappa$-$\gamma$ mechanism \citep{1981A&A...102..375D} and convective driving mechanism \citep{1991MNRAS.251..673B}. They provide a unique opportunity to explore the invisible interior of ELM WDs by analysing pulsation modes and matching them with theoretical models to obtain information about their chemical composition and structure, to determine their stellar parameters (mass, effective temperature, rotation period, etc.), and to test the scenarios of their formation by employing asteroseismic tools \citep{2008ARA&A..46..157W,2008PASP..120.1043F,2010A&ARv..18..471A}.

{\obj} (hereafter referred to as {\jid}) was originally discovered to be an ELMV by \citet{2013ApJ...765..102H}. Seven pulsation modes were detected on {\jid}, two of which with very short periods (107.56 and 134.275\,s), and are suspected to be {\pmd}s (pressure modes). Theoretical calculations have predicted long ago that {\pmd} pulsations can be excited on WDs \citep[see][for example]{1983ApJ...265..982S,1983ApJ...269..645S,1985ApJ...297..544H,1993ApJ...404..294K}. However, the typical periods of {\pmd}s are comparable to the dynamic time scale $\tau_\mathrm{dyn}\propto 1/\sqrt{\bar{\rho}}$, which are usually in the order of seconds or less on canonical C/O-core WDs, making it a challenge to detecte these pulsation modes. There have been several observational efforts to search for {\pmd} pulsations in WDs \citep[e.g.][]{1984AJ.....89.1732R,1994AJ....107..298K,2011A&A...525A..64S,2013A&A...558A..63C,2014MNRAS.437.1836K}, even with sufficient time resolution (hundreds of milliseconds or even shorter), but no convincing {\pmd} pulsation has ever been observed in a pulsating WD. It is most likely the very high surface gravity of a WD makes the amplitudes of {\pmd}s, which are primarily vertical displacements, too small to be detected. In fact, bona fide {\pmd} pulsations have been detected on ELM WD precursors \citep[pre-ELM WDs, see][for example]{2016ApJ...822L..27G}. These pre-ELM WDs are quite similar in structure to ELM WDs but have lower surface gravities. The discovery of short-period pulsation on {\jid} seems to be the first time that these elusive pulsations have been detected on a WD and need to be confirmed. It makes {\jid} an interesting object. On one hand, these {\pmd}s provide stronger constraints in a complementary way with the existing {\gmd}s to probe the interior of ELM WDs. On the other hand, we hope to provide a strong support for the confirmation of these suspected {\pmd}s through modeling and analysis. It is worth mentioning that some previous asteroseismic analysis of {\jid} have experimentally taken the potential {\pmd} pulsations into account. The work of \citet{2014A&A...569A.106C} is an earlier attempt to fit the two short-period pulsations in {\jid} with {\pmd}s. Their results suggest that if {\jid} has a lower mass ($M \approx 0.16$\,{\msun}) than estimated by spectroscopy \citep[$M=0.179\pm0.0012$\,{\msun},  derived using the models of][]{2013AA...557A..19A}, the short-period pulsations (107.56 and 134.275\,s) might be interpreted as low order {\pmd}s. Another work from \citet{2017AA...607A..33C} performed period fit to all the seven periods as one case of their analysis.

The paper is organized as follows. {\sn}~\ref{sec:ident} describes the process of identification of the observed  pulsation modes of {\jid}. A detailed introduction to the modeling for {\jid} is presented in {\sn}~\ref{sec:model}. {\sn}~\ref{subsec:theor} describes the theoretical models and {\sn}~\ref{subsec:aster} introduces the method for determining the best-fit model and discusses the results. The conclusions are given in {\sn}~\ref{sec:concl}.

\section{Identification of the observed pulsation modes} \label{sec:ident}

\begin{figure*}
\plotone{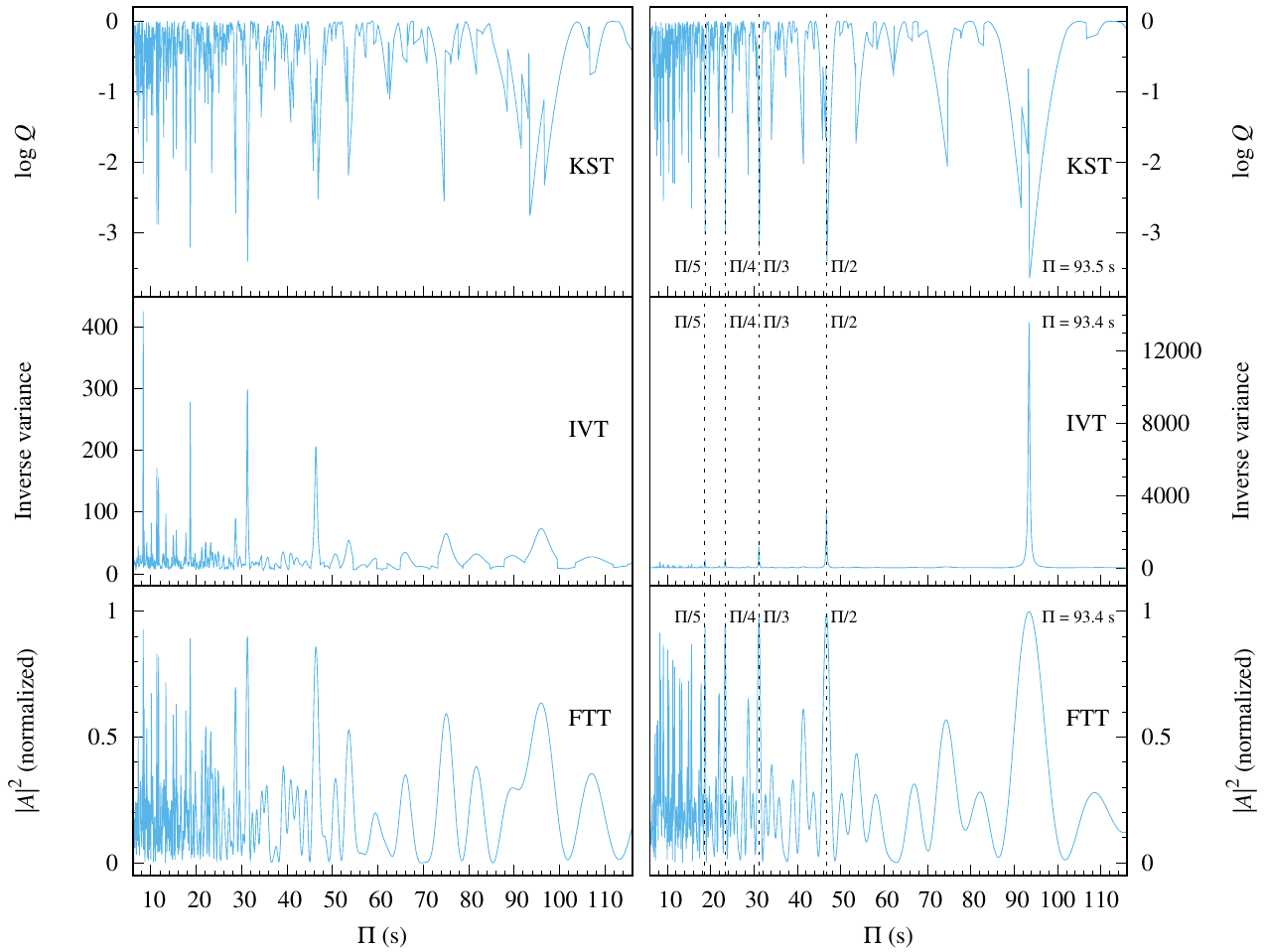}
\caption{The left panels display the results of three independent tests KST, IVT and FTT applied to periods of five {\gmd}s (1792.905, 1884.599, 2258.528, 2539.695 and 2855.728\,s) to search for equal period spacing. The right panels are the same as the left panels, but the results of eliminating the period 2855.728\,s. A clear and significant period spacing at $\Pi \approx 93.4$\,s is exhibited simultaneously in the three tests. The vertical dashed lines in right panels indicate the harmonics of the main period spacing ($\Pi/2$, $\Pi/3$, $\Pi/4$ and $\Pi/5$).}
\label{fig01}
\end{figure*}

An important step before doing asteroseismic analysis is to identify the spherical harmonic degree $\ell$ of the observed pulsation modes. The asymptotic analysis shows that the periods of high-order {\gmd}s with the same degree $\ell$ and consecutive radial order $n$ are expected to have approximately equal spacing \citep{1980ApJS...43..469T}, i.e.
\begin{equation}
\Pi_{\ell}=P_{\ell,\,n+1}-P_{\ell,\,n} \approx \frac{2\pi^2}{\sqrt{\ell(\ell+1)}}\left(\int_0^R\frac{N}{r}\ud r\right)^{-1},
\end{equation}
where $P_{\ell,\,n}$ represents the period of pulsation mode with $\ell$ and $n$, and $N$ is the Brunt-V{\"a}is{\"a}l{\"a} frequency. Finding out a constant period spacing from the observed data will help to determine the $\ell$ value. Three independent methods are used to search for the equal period spacing, namely the K-S test \citep[KST, see][]{1988IAUS..123..329K}, the inverse variance test \citep[IVT, see][]{1994MNRAS.270..222O}, and the Fourier transform test \citep[FTT, see][]{1997MNRAS.286..303H}. There are seven pulsation modes detected on {\jid} \citep{2013ApJ...765..102H}. We perform analysis on the five long periods (1792.905, 1884.599, 2258.528, 2539.695 and 2855.728\,s), excluding the two suspected {\pmd}s (107.56 and 134.275\,s). The results are shown in the left panel of {\fn}~\ref{fig01}. In the KST, the minimum value of $\log Q$ means a uniform period spacing is significant. In the IVT, the maximum value of the inverse variance represents a period spacing with high probability. Similarly, in the FTT, the peak in the square of amplitude ($|A|^2$) indicates a possible period spacing. As shown in the figure, all three tests fail to give a clear period spacing for the five periods. The reason is probably that these periods belong to different $\ell$, thus exhibiting inconsistent period spacing. If the period 2855.728\,s is eliminated, a clear and significant period spacing at $\Pi \approx 93.4$\,s appears simultaneously in the three tests, as shown in the right panel of {\fn}~\ref{fig01}. It provides reliable evidence that the four periods (1792.905, 1884.599, 2258.528, and 2539.695\,s) belong to the same $\ell$. According to \citet{2014A&A...569A.106C}, if $\Pi \approx 93.4$\,s is the asymptotic period spacing of $\ell=1$ {\gmd}s, the mass of the considered WD is expected to be 
approximately 0.16 to 0.17\,{\msun}, which is exactly consistent with the mass ($M \approx 0.17$\,{\msun}) estimated by \citet{2013ApJ...765..102H} from spectroscopy. The above four periods can therefore be identified as $\ell=1$ modes. The period 2855.728\,s should have a different $\ell$ value. It may be assumed to be an $\ell=2$ mode. Higher $\ell$ values are unrealistic, because geometrical cancellation makes $\ell>2$ modes hard to observe photometrically \citep{1982ApJ...259..219R}.

\section{Modeling} \label{sec:model}

\subsection{Theoretical models} \label{subsec:theor}

\begin{figure*}
\plotone{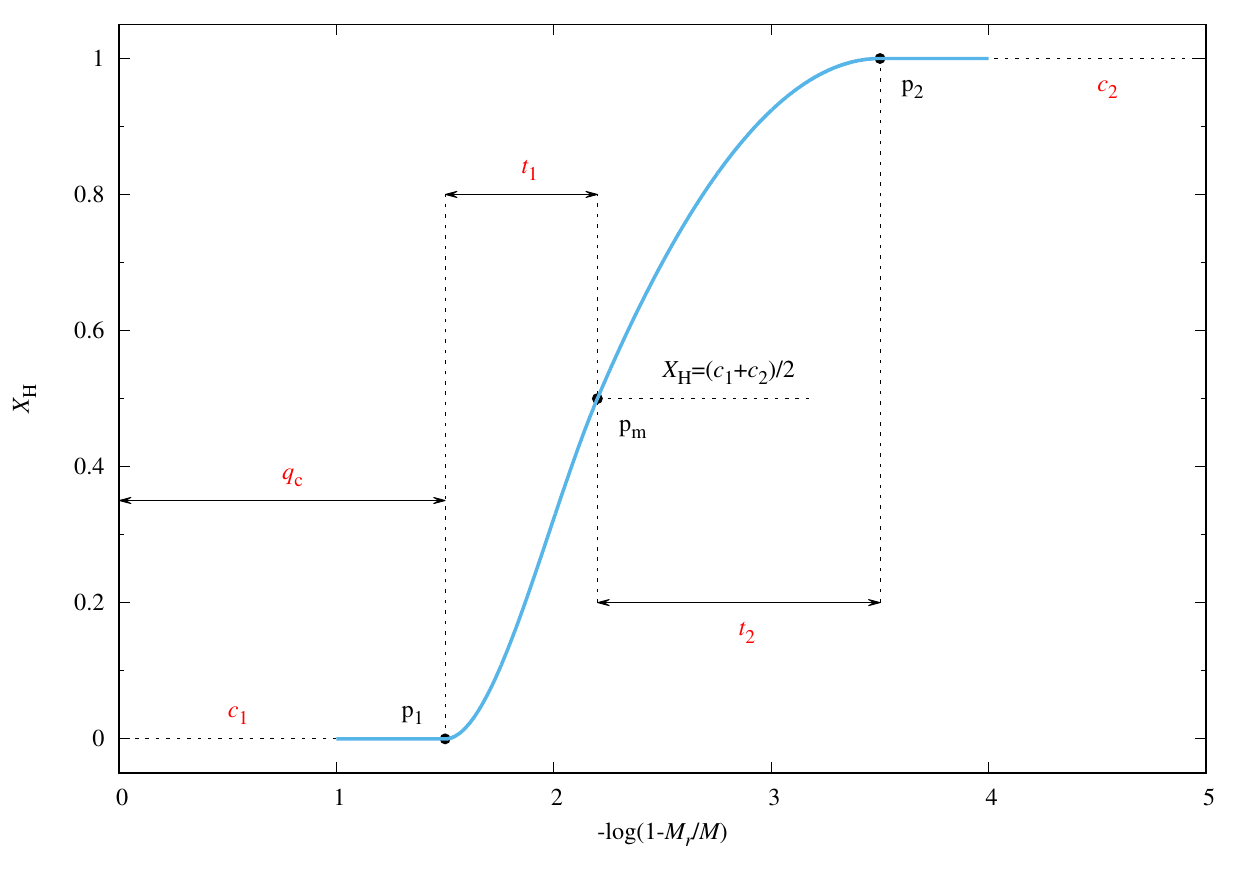}
\caption{Illustration of the parameterization scheme. Three black dots p$_1$, p$_\mathrm{m}$ and p$_2$ represent control points, which are defined by five parameters. All parameters are labeled in red on the diagram. The control points determine Akima splines (blue curve), which describe the H abundance profile of the H/He transition zone.}
\label{fig02}
\end{figure*}

In this work, we consider models with parametrized chemical profiles. The scheme of parametrizing chemical composition of WDs had been applied in several previous works \citep[see][for instance]{2006A&A...446..223P,2006A&A...453..219P,2008ApJ...675.1505B,2008MNRAS.385..430C,2009MNRAS.396.1709C,2016ApJS..223...10G,2017ApJ...834..136G,2018Natur.554...73G,2021ApJ...922..138G}. We adopt a scheme based on Akima splines \citep{1970JACM...17..589A} to imitate the H abundance profile of the H/He transition zone. A similar scheme was used in the works of \citet{2017ApJ...834..136G,2018Natur.554...73G} to describe the core chemical profiles of C/O-core WDs. It has been proven to improve the accuracy of modeling. The parameterization scheme is illustrated in {\fn}~\ref{fig02}. As shown in the figure, three control points (p$_1$, p$_\mathrm{m}$ and p$_2$) defined by five parameters are introduced to determine the H/He transition zone, and the parameters are 
\begin{enumerate}
\item $q_\mathrm{c}$ indicates the location of p$_1$, the core boundary,
\item $c_1$ represents the H abundance ($X_\mathrm{H}$) of the core,
\item $c_2$ represents $X_\mathrm{H}$ of the outer layer,
\item $t_1$ controls the scale of transition zone below the midpoint (p$_\mathrm{m}$), and
\item $t_2$ controls the scale of transition zone above p$_\mathrm{m}$,
\end{enumerate}
where p$_\mathrm{m}$ is defined as the position where $X_\mathrm{H}$ is always equal to $\frac{1}{2}\left(c_1+c_2\right)$.
Here we only consider H atmosphere WD models with pure He core. Therefore, the two parameters $c_1$ and $c_2$ are always fixed as constants, that is $c_1=0$ and $c_2=1$. There are only three free parameters $q_\mathrm{c}$, $t_1$ and $t_2$ then can be adjusted. The helium profile is given by $X_\mathrm{He}=1-X_\mathrm{H}$. 

The models for asteroseismic analysis are calculated using the Modules for Experiments in Stellar Astrophysics ({\mesa}, version number 12778), which is a suite of open source libraries for a wide range of applications in stellar astrophysics \citep[see][for details]{2011ApJS..192....3P,2013ApJS..208....4P,2015ApJS..220...15P,2018ApJS..234...34P,2019ApJS..243...10P}. The OPAL equation of state tables \citep{2002ApJ...576.1064R} are used in this work. Radiative opacities are adopted from OPAL tables \citep{1993ApJ...412..752I,1996ApJ...464..943I} in the high-temperature region and tables of \citet{2005ApJ...623..585F} in the low-temperature region, the latter include contributions from molecules and grains. Conductive opacities are taken from \citet{2007ApJ...661.1094C}. The nuclear reaction network used is the {\mesa} default {\tt\string basic.net} that contains eight isotopes: $^1$H, $^3$He, $^4$He, $^{12}$C, $^{14}$N, $^{16}$O, $^{20}$Ne, and $^{24}$Mg. We set the metal composition as the abundance ratio of GS98 \citep{1998SSRv...85..161G}. The mixing length theory (MLT) formulation of \citet{1971A&A....12...21B} is employed to deal with convection. The mixing length parameter is set to $\alpha_\mathrm{MLT}=2$ ({\mesa} default setting) during the evolution of WD progenitor, and $\alpha_\mathrm{MLT}=0.6$ during the evolution of WD, which refers to the value recommended by \citet{1995ApJ...449..258B}. We include element diffusion in our models using the scheme described by \citet{1994ApJ...421..828T}, but ignore convective overshoot and rotational mixing. It is worth mentioning that rotation plays a critical role in determining the structure of ELM WD as well as diffusion \citep{2016A&A...595A..35I}. Ignoring effects such as rotation would lead to inaccurate models. However, introducing parameterized chemical profiles into WD models is expected to reduce some inaccuracies.

\begin{figure*}
\plotone{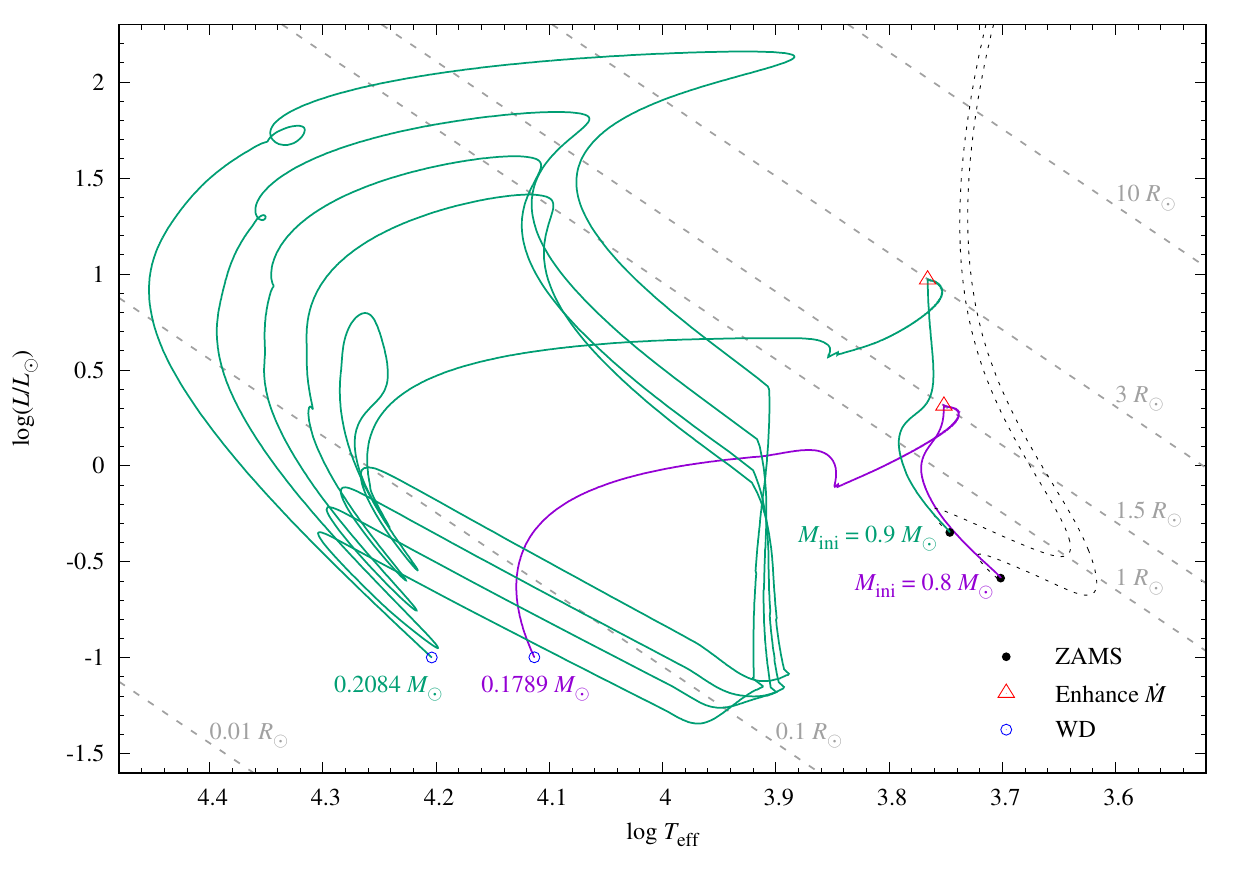}
\caption{The H-R diagram shows the formation of our He WD models. The black dotted lines represent the evolutionary tracks of the pre-main sequence stages and the black points indicate the locations of the zero-age main sequence (ZAMS). The red triangles in the figure indicate the stages where the mass loss rate ($\dot{M}$) are enhanced. It shows that the model with $M_\mathrm{ini}=0.9$\,{\msun} experiences multiple diffusion-induced CNO flashes and finally formes a WD with $M=0.2084$\,{\msun}. The other model with $M_\mathrm{ini}=0.8$\,{\msun} do not experience a similar process, but directly formes a WD with $M=0.1789$\,{\msun}. The grey dashed lines in the background represent grid lines corresponding to different {\rsun} values.}
\label{fig03}
\end{figure*}

The He WD models are generated from the zero-age main sequence (ZAMS) progenitors through stellar evolution. They start with two models with initial masses ($M_\mathrm{ini}$) of 0.8 and 0.9\,{\msun}, respectively. All models are assumed to have initial chemical compositions of $(X, Y, Z)=(0.7, 0.28, 0.02)$. We evolve the model with $M_\mathrm{ini}=0.8$\,{\msun} until its radius exceeds 1.5\,{\rsun}, and then the mass loss rate is enhanced, allowing a large amount of the envelope to be stripped off before the He flash occurs. For the model with $M_\mathrm{ini}=0.9$\,{\msun}, this process is started when its radius exceeds 3\,{\rsun}. In the subsequent evolution, the model with $M_\mathrm{ini}=0.9$\,{\msun} goes through repeated diffusion-induced CNO flashes \citep[e.g.][]{2001MNRAS.323..471A,2007MNRAS.382..779P,2013AA...557A..19A,2016A&A...595A..35I} and then arrive at the WD cooling track. The evolution is stopped when $\log\left(L/L_\sun\right)<-1$ and an initial WD model with mass of 0.2084\,{\msun} is obtained (hereafter referred to as {\wdb}). However, the model with $M_\mathrm{ini}=0.8$\,{\msun} do not experience a similar process of CNO flashes, but directly form a WD with a lower mass of 0.1789\,{\msun} (hereafter referred to as {\wda}). {\fn}~\ref{fig03} shows the evolutionary tracks of the two progenitors in the Hertzsprung-Russell (H-R) diagram. An example of the {\tt\string inlist} options for the model with $M_\mathrm{ini}=0.8$\,{\msun} is listed in {\an}~\ref{app01} for reference (the options for the model with $M_\mathrm{ini}=0.9$\,{\msun} are similar). In addition, we made some modifications to the source file {\extf} to add custom options. The contents of the modified {\extf} file are shown in {\an}~\ref{app02}.

The parameterized H/He profiles are calculated separately and incorporated into {\mesa} using the {\tt\string relax{\_}initial{\_}composition} option to get models with different chemical profiles. An example of the {\tt\string inlist} for relaxing the composition of a model is shown in {\an}~\ref{app03}. The range and step of the profile parameters, $q_\mathrm{c}$, $t_1$ and $t_2$, adopted in different grids are listed in {\tn}~\ref{tab01}. The mass of a certain model with given chemical profiles is rescaled without changing composition by setting the {\tt\string relax{\_}mass{\_}scale} option to obtain models with different masses. We relax {\wda} to create models with masses of $0.15 \leqslant M < 0.18$\,{\msun} and relax {\wdb} to create WD models with masses of $0.18 \leqslant M \leqslant 0.2$. The model grid used for this analysis covers the range in $M$ from 0.15 to 0.2\,{\msun} in step of 0.001 (coarse grid) and 0.0001\,{\msun} (fine grid), and ${\teff}$ from 7000 to 12000\,K. An example of the {\tt\string inlist} for relaxing the mass of a model and cooling it are shown in {\an}~\ref{app04}. The stellar oscillation code {\gyre} \citep{2013MNRAS.435.3406T} is adopted to calculate the eigenfrequencies and eigenfunctions of the oscillation modes of each model in the grid. The Brunt-V{\"a}is{\"a}l{\"a} frequency from the model is smoothed by the built-in weighted smoothing in {\mesa} using two cells on either side ({\mesa} default setting), in order to reduce the influence of numerical noise on theoretical frequencies.

\begin{deluxetable*}{ccccccccccccc}
\tablecaption{The minimum (Min.), maximum (Max.) and step values of the profile parameters, $q_\mathrm{c}$, $t_1$ and $t_2$, adopted in four grids with different resolutions.}
\label{tab01}
\tablehead{
\nocolhead{} && \multicolumn{3}{c}{$q_\mathrm{c}$} && \multicolumn{3}{c}{$t_1$} && \multicolumn{3}{c}{$t_2$} \\
\cline{3-5} \cline{7-9} \cline{11-13}
\nocolhead{} && \colhead{Min.} & \colhead{Max.} & \colhead{Step} && \colhead{Min.} & \colhead{Max.} & \colhead{Step} && \colhead{Min.} & \colhead{Max.} & \colhead{Step}
} 
\startdata
Grid 1 && 0.30 & 0.80 & 0.1   && 0.50 & 1.50 & 0.1   && 0.50 & 1.50 & 0.1 \\
Grid 2 && 0.30 & 0.50 & 0.05  && 0.60 & 0.80 & 0.05  && 1.20 & 1.40 & 0.05 \\
Grid 3 && 0.35 & 0.45 & 0.01  && 0.65 & 0.75 & 0.01  && 1.30 & 1.40 & 0.01 \\
Grid 4 && 0.39 & 0.41 & 0.001 && 0.68 & 0.70 & 0.001 && 1.33 & 1.35 & 0.001 \\
\enddata
\end{deluxetable*}

\subsection{Asteroseismic analysis} \label{subsec:aster}

The analysis is based on the forward modeling approach, which involves finding the best match between a set of oscillation frequencies detected in a given star and the frequencies calculated from the models. The goodness of matching for each model is quantified by a merit function defined as
\begin{equation}
S^2\equiv\frac{1}{N}\sum_{i=1}^{N}\left[f_\mathrm{O}(i)-f_\mathrm{T}\right]^2,
\end{equation}
where $f_\mathrm{O}(i)$ is the frequency of the $i$-th observed mode, $f_\mathrm{T}$ is the theoretical frequency which is the closest one to $f_\mathrm{O}(i)$, and $N=7$ is the total number of observed modes. The model with minimum $S^2$ is considered to be the best-fit model. In Section \ref{sec:ident}, four of the observed {\gmd}s (557.7542, 530.6170, 442.7662 and 393.7480\,{\uhz}) are identified to be $\ell=1$ modes and the other one (350.1734\,{\uhz}) to be $\ell=2$ mode. Each of them only needs to match the theoretical frequency of the corresponding $\ell$ value. As for the two suspected {\pmd}s (7447.388 and 9297.4\,{\uhz}), due to the lack of clear identification of their $\ell$ values, each of them will match the theoretical frequency of $\ell=0, 1, \text{and }2$, respectively. ELM WDs are expected to be relatively rapidly rotating \citep[e.g.][]{2016A&A...595A..35I}. Thus, rotational splittings are expected to be observed in all the non-radial pulsation modes. However, no triplet or quintet produced by rotational splitting was observed among the seven observed modes. We assumed implicitly that all observed modes are $m=0$, because there is not enough evidence to confirm their $m$ values. Therefore, we only explore the theoretical modes of $m=0$, and do not consider rotational multiplets. However, this assumption is somewhat suspicious. We should be aware that this would bring some uncertainties to the results.

\begin{figure*}
\plotone{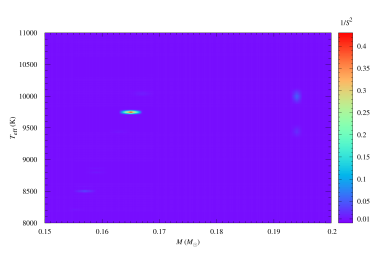}
\caption{The heat map shows the projection of $1/S^{2}$ on the $M$ versus ${\teff}$ plane. The best-fit model is located at $M=0.1650$\,{\msun} and ${\teff}=9750$\,K, which corresponds to the maximum of $1/S^{2} \approx 0.423$ ($S^2 \approx 2.366$).}
\label{fig04}
\end{figure*}

\begin{figure*}
\plotone{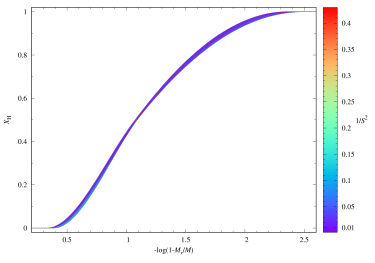}
\caption{The variation of $1/S^{2}$ when $M$ and ${\teff}$ are fixed and the H chemical profile of the best-fit model is fine tuned. The value of $1/S^2$ is indicated by the color scale.}
\label{fig05}
\end{figure*}

We find out the best-fit model with a minimum $S^2$ of 2.366. The mass and effective temperature of the best-fit model are $M=0.1650$\,{\msun} and ${\teff}=9750$\,K. The parameters characterizing its H chemical profile are $q_\mathrm{c}=0.399$, $t_1=0.685$ and $t_2=1.343$, which correspond to an H envelope mass of $\log\left(M_\mathrm{H}/M\right)=-1.45$. {\fn}~\ref{fig04} shows the projection of the inverse of the merit function, $1/S^2$, on the $M$ versus ${\teff}$ plane with the chemical profiles to be fixed as the best-fit solution. {\fn}~\ref{fig05} shows $1/S^2$ varies with the H chemical profile when $M$ and ${\teff}$ are fixed, where the H chemical profile of the best-fit model is fine tuned and the value of $1/S^2$ is indicated by the color scale.

\begin{figure*}
\plotone{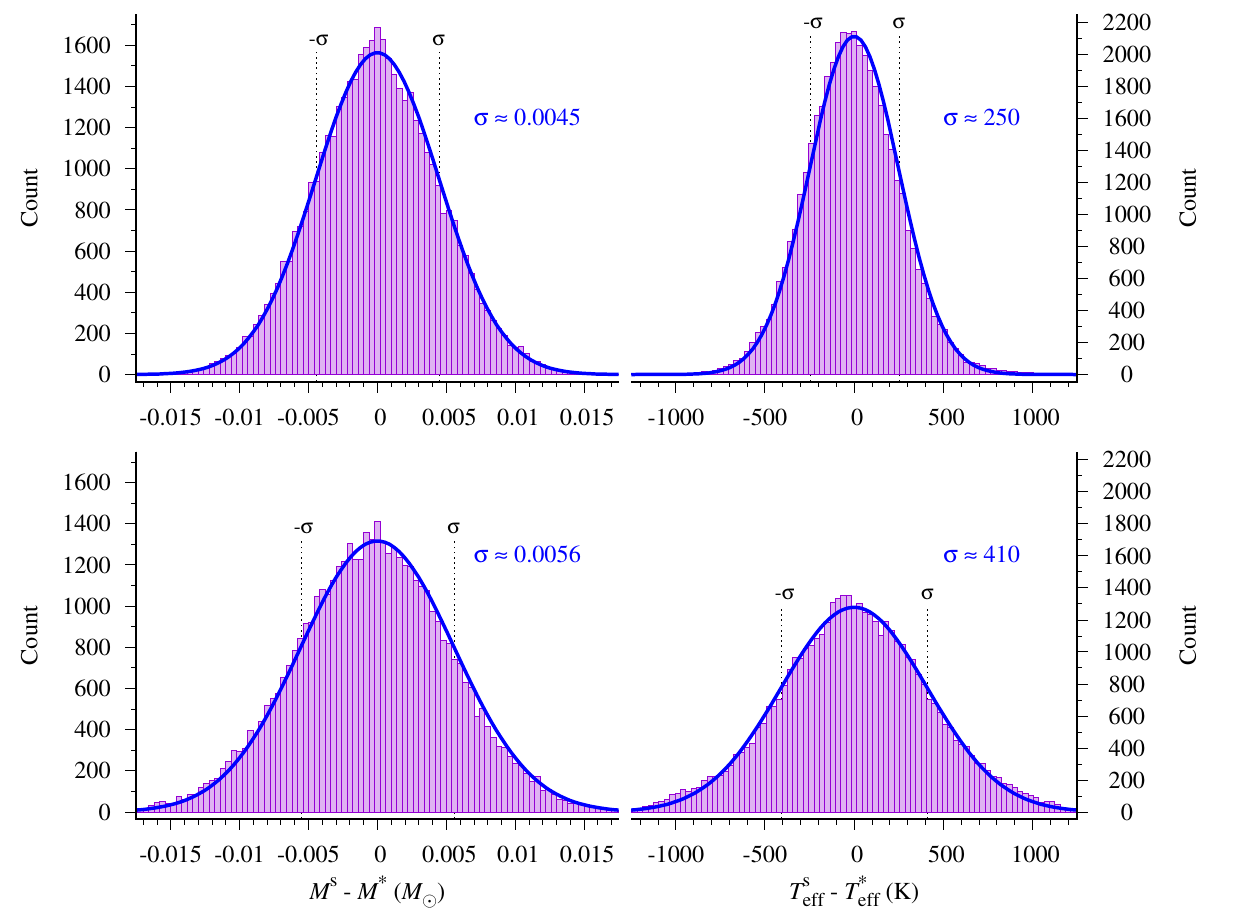}
\caption{Histograms of the distribution of simulated parameters $M^\mathrm{s}$ and $\teff^\mathrm{s}$ around the best-fit parameters $M^{\ast}=0.1650$\,{\msun} and $\teff^{\ast}=9750$\,K. The blue curves are results of fitting the histograms with Gaussian function, from which the standard deviations ($\sigma$) are obtained. The upper panels show the results for the case where the simulated random noises are equivalent to the observational uncertainties, which give $\sigma \approx 0.0045$\,{\msun} for mass and $\sigma \approx 250$\,K for effective temperature. The lower panels show the results for the case where the simulated random noises correspond to the differences between the observed and theoretical frequencies. The results for this case are $\sigma \approx 0.0056$\,{\msun} for mass and $\sigma \approx 410$\,K for effective temperature.}
\label{fig06}
\end{figure*}

We use the bootstrap method \citep{10.1214/aos/1176344552} to estimate the uncertainties of $M$ and ${\teff}$ of the best-fit model. The bootstrap is a Monte Carlo simulation based on a large number of synthetic data sets that sampled from the original data set. The size of each synthetic data set must be equal to the size of the original data set. First, we sample seven data points at a time with replacement from the seven observed frequencies to form a new data set (called bootstrap sample). To account for the perturbation of each frequency by its observational uncertainty ($\sigma_{\rm f}$), a normally distributed random noise with zero mean and standard deviation $\sigma_{\rm f}$ is added to each sample point. Then the bootstrap sample is taken to match the models to obtain parameters with the minimum $S^2$. The fitted parameters obtained in this way are expressed as simulated parameters ($M^\mathrm{s}$ and $\teff^\mathrm{s}$). This procedure is repeated 50000 times in this work. The distribution of these $M^\mathrm{s}$ and $\teff^\mathrm{s}$ around the best-fit parameters $M^{\ast}=0.1650$\,{\msun} and $\teff^{\ast}=9750$\,K is then analyzed. Histograms of the distribution of $M^\mathrm{s}-M^{\ast}$ and $\teff^\mathrm{s}-\teff^{\ast}$ are shown in the upper panels of {\fn}~\ref{fig06}. We fit each histogram with a Gaussian function and get the standard deviation ($\sigma$) from the fitting curve. It gives $\sigma \approx 0.0045$\,{\msun} for mass and $\sigma \approx 250$\,K for effective temperature. However, $\sigma_{\rm f}$ is generally less than $0.002$\,{\uhz}, which is several orders of magnitude smaller than the frequency difference between the observed and theoretical frequencies ($|f_\mathrm{T}-f_\mathrm{O}|$, see {\tn}~\ref{tab03}). In order to accommodate the large mismatch between the observed modes compared to the theoretical ones, we try to introduce normally distributed random noise with standard deviation $|f_\mathrm{T}-f_\mathrm{O}|$ to perform bootstrapping analysis again. The distributions of $M^\mathrm{s}-M^{\ast}$ and $\teff^\mathrm{s}-\teff^{\ast}$ along with the corresponding fitting curve are shown in the lower panels of {\fn}~\ref{fig06}. The results for this case are $\sigma \approx 0.0056$\,{\msun} for mass and $\sigma \approx 410$\,K for effective temperature.

We can also estimate the distance to {\jid} by using the luminosity of the best-fit model ($L=0.067624\,L_{\sun}$) and the Gaia apparent magnitude of {\jid} ($G=16.335$\,mag), which is similar to what \citet{2021ApJ...922..220L} has done recently. We use the 3D reddening map of \citet{2018MNRAS.478..651G} and apply the $R_\mathrm{V}=3.1$ reddening law of \citet{1999PASP..111...63F} to obtain the extinction of $A_\mathrm{V}=0.073$. We then use the bolometric correction of \citet{2016ApJ...823..102C} for a star with $\teff=9750$\,K, $\log g=6.0$ and $A_\mathrm{V}=0.1$ to determine the seismic distance of 434.3\,pc. However, this value deviates significantly from the parallax distance determined from Gaia eDR3 of $361.4^{\,+\,6.5}_{\,-\,7.7}$\,pc \citep{2021AJ....161..147B}, which suggests that the luminosity of our model is greater than that measured by Gaia parallax. If we assume that the seismic distance of our model agrees with the parallax distance at the 1-$\sigma$ level, the seismic distance should be estimated to be $434.3\pm66.4$\,pc. In order to be consistent with the uncertainty of this distance, the luminosity of our model should lie in the range of 0.048526 to 0.089881\,$L_{\sun}$. We try to use the variance of luminosity to estimate the uncertainties of our model parameters, which gives $\sigma \approx 0.0137$\,{\msun} for mass and $\sigma \approx 560$\,K for effective temperature. It suggests that the uncertainties of mass and effective temperature derived from bootstrap simulations are somewhat underestimated. Based on the above analysis, we will choose the uncertainties estimated by the last scheme as the final results, i.e. $M=0.1650\pm0.0137$\,{\msun} and ${\teff}=9750\pm560$\,K, which are considered more reliable.

The atmospheric parameters of {\jid} from spectroscopic measurement were previously published by \citet{2012ApJ...744..142B} and \citet {2013ApJ...765..102H}. The current values include corrections of \citet{2015ApJ...809..148T} for the 3D treatment of convection, i.e. $\teff=9240\pm140$\,K and $\log g=6.17\pm0.06$, which corresponds to a stellar mass of 0.169 \,{\msun}. The stellar mass and effective temperature of our best-fit model are consistent with the spectroscopic results and the asteroseismic solutions ($M=0.1612$\,{\msun} and ${\teff}=9301$\,K) of \citet{2018AA...620A.196C} within 1-$\sigma$ uncertainty. The hydrogen-layer mass ($M_\mathrm{H}/M$) of {\jid} derived from the present work are slightly greater than the asteroseismic solutions of \citet{2017AA...607A..33C,2018AA...620A.196C}, i.e. $\log\left(M_\mathrm{H}/M\right)=-2.45$ and $\log\left(M_\mathrm{H}/M\right)=-1.76$, respectively. It is worth mentioning that the results of \citet{2018AA...620A.196C} was based on the five observed {\gmd}s, but our analysis takes the whole set of (seven) frequencies into account. Moreover, the differences between the two results may reflect some differences between the theoretical models of \citet{2017AA...607A..33C,2018AA...620A.196C} and ours. Moreover, the mass of our best-fit is in agreement with the previous estimates of $M \approx 0.17$\,{\msun} by \citet{2013ApJ...765..102H} and $M = 0.169$\,{\msun} by \citet{2015ApJ...809..148T}. It is also consistent with the suggestion of \citet{2014A&A...569A.106C} that the two short-period pulsations might be caused by low-order {\pmd}s if the mass of {\jid} is $M \approx 0.16$\,{\msun}. A comparison between the stellar parameters obtained in this work and the results of previous work is listed in {\tn}~\ref{tab02}.

\begin{deluxetable*}{ccccc}
\tablecaption{Comparison between the stellar parameters obtained in this work and the previous results.}
\label{tab02}
\tablehead{
\colhead{$M$\,({\msun})} & \colhead{$\teff$\,(K)} & \colhead{$\log g$\,(cgs)} & \colhead{$\log\left(M_\mathrm{H}/M\right)$} & \colhead{Work}}
\startdata
$0.1650\pm0.0137$ & $9750\pm560$         & $5.73\pm0.15$         & $-1.45$ & This \\
$0.16^{\,a}$      & $(9400\pm490)^{\,b}$ & $(5.81\pm0.12)^{\,b}$ & -       & \citet{2012ApJ...744..142B} \\
$0.17^{\,a}$      & $(9590\pm140)^{\,b}$ & $(6.36\pm0.06)^{\,b}$ & -       & \citet{2013ApJ...765..102H} \\
$0.179\pm0.0012$  & $(9590\pm140)^{\,b}$ & $(6.36\pm0.06)^{\,b}$ & -       & \citet{2013AA...557A..19A} \\
$0.169$           & $9240\pm140$         & $6.17\pm0.06$         & -       & \citet{2015ApJ...809..148T} \\
$0.2389$          & $9300$               & $6.9215$              & $-2.45$ & \citet{2017AA...607A..33C} \\
$0.1612$          & $9301$               & $5.9695$              & $-1.76$ & \citet{2018AA...620A.196C} \\
\enddata
\tablenotetext{}{$^{a}$ Estimated using the models of \citet{2007MNRAS.382..779P}.}
\tablenotetext{}{$^{b}$ Preliminary spectroscopic results without 3D corrections.}
\end{deluxetable*}

A part of the theoretical frequencies ($f_\mathrm{T}$) of the best-fit model are listed in {\tn}~\ref{tab03} along with their spherical harmonic degree $\ell$ and radial order $n$. The observed frequencies ($f_\mathrm{O}$) and the corresponding period and amplitude values are listed beside the matched theoretical frequencies. The absolute frequency differences $|f_\mathrm{T}-f_\mathrm{O}|$ are also listed in the table. We hence identify the $\ell$ and $n$ values of the seven observed modes. Four observed periods 1792.905, 1884.599, 2258.528, and 2539.695\,s (frequencies 557.7542, 530.6170, 442.7662 and 393.7480\,{\uhz}), have been identified as $\ell=1$ {\gmd} in {\sn} \ref{sec:ident}, and their $n$ values are determined as 17, 18, 22 and 25 by comparing them to the theoretical modes. The period 2855.728\,s (frequency 350.1734\,{\uhz}) is identified to be a {\gmd} with $\ell=2$ and $n=50$. The two suspected {\pmd}, 107.56 and 134.275\,s (9297.4 and 7447.388 \,{\uhz}), are identified as $\ell=2$, $n=2$ and $\ell=1$, $n=2$, respectively. It is worth noting that the periods calculated by the best-fit model, 107.567 and 134.258\,s, are in excellent agreement with these two observations (107.56 and 134.275\,s).

\begin{deluxetable*}{ccccccccc}
\tablecaption{Comparison of the theoretical and observed frequencies of the best-fit model. The theoretical frequencies $f_\mathrm{T}$ are listed along with their spherical harmonic degree $\ell$ and radial order $n$. The seven observed frequencies $f_\mathrm{O}$ and the corresponding period and amplitude values are listed beside the matched theoretical frequencies. The absolute frequency differences $|f_\mathrm{T}-f_\mathrm{O}|$ are also listed.}
\label{tab03}
\tablehead{
\nocolhead{} & \multicolumn{4}{c}{Theoretical} & \nocolhead{} & \multicolumn{3}{c}{Observed \citep{2013ApJ...765..102H}} \\
\cline{2-5} \cline{7-9}
\nocolhead{} & \colhead{$\ell$} & \colhead{$n$} & \colhead{Period} & \colhead{Frequency} & \colhead{$|f_\mathrm{T}-f_\mathrm{O}|$} & \colhead{Frequency} & \colhead{Period} & \colhead{Amplitude} \\
\nocolhead{} & \nocolhead{} & \nocolhead{} & \colhead{({s})} & \colhead{({\uhz})} & \colhead{({\uhz})} & \colhead{({\uhz})} &\colhead{(s)} & \colhead{(mma)}
}
\startdata
\multirow{2}{*}{{\pmd}s} & 1 &  2 & 134.258 & 7448.3354 & 0.9474 & $7447.388\pm0.010$ & 134.275 & 0.44 \\
& 2 &  2 & 107.567 & 9296.5663 & 0.8337 & $9297.4\pm3.6$ & 107.56 & 0.38 \\
\cline{2-9}
\multirow{5}{*}{{\gmd}s} & 1 & $-17$ & 1787.030 & 559.5877 & 1.8335 & $557.7542\pm0.0017$ & 1792.905 & 3.31 \\
& 1 & $-18$ & 1875.696 & 533.1355 & 2.5185 & $530.6170\pm0.0011$ & 1884.599 & 4.73 \\
& 1 & $-22$ & 2262.705 & 441.9489 & 0.8173 & $442.7662\pm0.0007$ & 2258.528 & 7.49 \\
& 1 & $-25$ & 2551.858 & 391.8713 & 1.8767 & $393.7480\pm0.0007$ & 2539.695 & 6.77 \\
& 2 & $-50$ & 2864.230 & 349.1340 & 1.0394 & $350.1734\pm0.0013$ & 2855.728 & 3.63 \\
\enddata
\end{deluxetable*}

\section{Conclusions} \label{sec:concl}

{\obj} is the second ELMV discovered. Two short-period pulsations were detected on this star, which means that {\pmd} pulsations may be observed on WDs for the first time. However, the reality of these two {\pmd}s has not been definitively confirmed
observationally. In this work, we perform a detailed asteroseismic analysis on {\jid}, taking the whole set of pulsation modes into account. For the five {\gmd}s, three independent methods are used to help us identify that four of them are $\ell=1$ modes and the other is $\ell=2$ mode. On this basis, we make asteroseismic modeling for the star, in which the H chemical profile is taken as a variable. The main parameters are determined from the best-fit model and the H/He chemical profiles are defined. The effective temperature and mass of {\jid} determined from our model are in good agreement with the parameters derived from spectroscopy and also are compatible with the results of other asteroseismic analysis. It is important that the two suspected {\pmd}s, 107.56 and 134.275\,s (9297.4 and 7447.388\,{\uhz}), are well represented in our best-fit model. Both the stellar parameters and the pulsation frequencies are in good agreement with the observations. We have found a model with {\pmd} pulsations consistent with the periods observed on {\jid}, which provides theoretical support for the reality of these two {\pmd}s.

Further observations are needed in the future. It can be expected that, with the development of observations on more ELMVs, it will be possible to discover more {\pmd} pulsations.

\begin{acknowledgments}

We would like to thank an anonymous referee for reviewing and offering valuable comments, which greatly help us to improve the manuscript. This research is supported in part by the National Key R\&D Program of China (Grant No. 2021YFA1600400/2021YFA1600402), by the B-type Strategic Priority Program No. XDB41000000 funded by the Chinese Academy of Sciences (CAS), and by the National Natural Science Foundation of China (NSFC) under grants 12133011 and 11833006. This work makes use of the ``PHOENIX Supercomputing Platform'' jointly operated by the Binary Population Synthesis Group and the Stellar Astrophysics Group at Yunnan Observatories, CAS.

\end{acknowledgments}

\appendix

\section{An example of the {\tt\string inlist} for generating He WD from ZAMS} \label{app01}

\begin{verbatim}
&star_job
      show_log_description_at_start = .false.
      create_pre_main_sequence_model = .true.
      set_uniform_initial_composition = .true.
        initial_h1 = 0.7
        initial_h2 = 0.0
        initial_he3 = 0.0
        initial_he4 = 0.28
        initial_zfracs = 3      !< GS98_zfracs = 3 >
      kappa_file_prefix = `gs98'
      eos_file_prefix = `mesa'
      save_photo_when_terminate = .false.
      save_model_when_terminate = .true.
      save_model_filename = `he_wd.mod'
      write_profile_when_terminate = .true.
      filename_for_profile_when_terminate = `he_wd_xxxx.dat'
/ ! end of star_job namelist

&controls
      initial_mass = 0.8
      initial_z = 0.02
      atm_option = `T_tau'
      atm_T_tau_relation = `Eddington'
      atm_build_tau_outer = 1.0d-6    
      MLT_option = `ML2'
      mixing_length_alpha = 2.0
      cool_wind_RGB_scheme = `Reimers'
      Reimers_scaling_factor = 0.5
      do_element_diffusion = .true.
!<--------------------------------------------------------------
! solve the system of equations described by Thoul et al. (1994)      
      diffusion_use_cgs_solver = .false.
!-------------------------------------------------------------->
      diffusion_min_dq_at_surface = 1.0d-3
      profile_interval = -1
      history_interval = 1
      terminal_interval = 1
      write_header_frequency = 100
      write_profiles_flag = .false.
      remove_H_wind_H_mass_limit = 1.0d-2
!<---------------------------------------------------------
!  custom option
!  enhance the mass loss rate when R/Rsun > this value
      x_ctrl(1) = 1.5 
!--------------------------------------------------------->
!<------------------------------------------------------------
!  custom stopping condition
!  stop when log(L/Lsun) < this value                           
      x_ctrl(2) = -1.d0
!------------------------------------------------------------>
/ ! end of controls namelist
\end{verbatim}

\section{Modifications to the source file `{\tt\string run\_star\_extras.f}'} \label{app02}

\begin{verbatim}
module run_star_extras
  use star_lib
  use star_def
  use const_def
  use math_lib    
  contains

  subroutine extras_controls(id, ierr)
    integer, intent(in) :: id
    integer, intent(out) :: ierr
    type (star_info), pointer :: s
    ierr = 0
    call star_ptr(id, s, ierr)
    if (ierr /= 0) return
    s%extras_check_model => extras_check_model
    s%job%warn_run_star_extras = .false.    
  end subroutine extras_controls

  integer function extras_check_model(id)
    integer, intent(in) :: id
    integer :: ierr
    type (star_info), pointer :: s
    ierr = 0
    call star_ptr(id, s, ierr)
    if (ierr /= 0) return
    extras_check_model = keep_going
    if (s%center_h1 < 1.0d-10 .and. s%log_surface_radius > log10(s%x_ctrl(1))) &
      s%remove_H_wind_mdot = 1.0d-7
    if (s%center_gamma > 1.0d0 .and. s%log_L_surf < s%x_ctrl(2)) then
      extras_check_model = terminate
      write(*, *) `have reached desired conditions'
      return
    end if
    if (extras_check_model == terminate) &
      s%termination_code = t_extras_check_model
  end function extras_check_model
end module run_star_extras
\end{verbatim}

\section{An example of the {\tt\string inlist} for relaxing the composition of a model} \label{app03}

\begin{verbatim}
&star_job
      show_log_description_at_start = .false.
      load_saved_model = .true.
      saved_model_name = `he_wd_xxxx.dat'
      kappa_file_prefix = `gs98'
      eos_file_prefix = `mesa'
      relax_initial_composition = .true.
      relax_composition_filename = `comp_xxxx.dat'
      num_steps_to_relax_composition = 100
      save_model_when_terminate = .true.
      save_model_filename = `wd_comp_xxxx.mod'
      steps_to_take_before_terminate = 10
/ ! end of star_job namelist

&controls   
      atm_option = `T_tau'
      atm_T_tau_relation = `Eddington'
      atm_build_tau_outer = 1.0d-6
      mesh_delta_coeff = 0.2
      dxdt_nuc_factor = 0.0
      eps_nuc_factor = 0.0
      mix_factor = 0.0  
      profile_interval = -1
      history_interval = -1
      terminal_interval = 1
      write_header_frequency = 100
      max_years_for_timestep = 1.0
/ ! end of controls namelist
\end{verbatim}

\section{An example of the {\tt\string inlist} for relaxing the mass of a model and cooling it} \label{app04}

\begin{verbatim}
&star_job
      show_log_description_at_start = .false.
      load_saved_model = .true.
      saved_model_name = `wd_comp_xxxx.mod'
      kappa_file_prefix = `gs98'
      eos_file_prefix = `mesa'
      relax_mass_scale = .true.
      new_mass = 0.16      !< new mass = 0.16 Msun, for example > 
      dlgm_per_step = 1.0d-3
      change_mass_years_for_dt = 1.0
      set_initial_dt = .true.
      years_for_initial_dt = 1.0
/ ! end of star_job namelist

&controls
      atm_option = `T_tau'
      atm_T_tau_relation = `Eddington'
      atm_build_tau_outer = 1.0d-6     
      add_atmosphere_to_pulse_data = .true.
      MLT_option = `ML2'
      mixing_length_alpha = 0.6      !< recommended by Bergeron et al. (1995) >
      mesh_delta_coeff = 0.2
      profile_interval = 1
      history_interval = 1
      terminal_interval = 1
      write_header_frequency = 100
      max_num_profile_models = -1
      profile_header_include_sys_details = .false.
      write_pulse_data_with_profile = .true.
      pulse_data_format = `GYRE'
      Teff_lower_limit = 7.0d3
/ ! end of controls namelist
\end{verbatim}

\end{CJK*}
\end{document}